\documentstyle[prl,aps,floats,twocolumn]{revtex}

\input{epsf}  %included so that figure could be embedded

\newcommand{\be}{\begin{equation}}
\newcommand{\ee}{\end{equation}}
\newcommand{\ba}{\begin{array}}
\newcommand{\ea}{\end{array}}
\newcommand{\bea}{\begin{eqnarray}}
\newcommand{\eea}{\end{eqnarray}}
\newcommand{\bml}{\begin{mathletters}}
\newcommand{\eml}{\end{mathletters}}
\newcommand{\bfi}{\begin{figure}}
\newcommand{\efi}{\end{figure}}

\begin{document}

\draft 

\title{Zero-Temperature Casimir Fluctuations\\
and the Limits of Force Microscope Sensitivity}
\author{J. A. Sidles \protect{\cite{pr}}}
\address{University of Washington School of Medicine}
\date{October 16, 1997
\protect\vspace{-0.3in}} % hack to fit on four pages

\maketitle
\widetext         % hack so the abstract appears nicely
\begin{abstract}
\leftskip 54.8pt  % hack so the abstract appears nicely
\rightskip 54.8pt % hack so the abstract appears nicely
It is predicted that in force microscopy the quantum fluctuations 
responsible for the Casimir force can be directly observed as 
temperature-independent force fluctuations having spectral density $9 
\pi/(40 \ln(4/e))\,\hbar\,|\delta k|$, where $\hbar$ is Planck's 
constant and $\delta k$ is the observed change in spring constant as 
the microscope tip approaches a sample.  For typical operating 
parameters the predicted force noise is of order $10^{-18}$ Newton in 
one Hertz of bandwidth.  The Second Law is respected via the 
fluctuation-dissipation theorem.  For small tip-sample separations the 
cantilever damping is predicted to increase as temperature is reduced, 
a behavior that is reminiscent of the Kondo effect.
\end{abstract} 
\pacs{    
\leftskip 54.8pt % hack so the pacs numbers appear nicely
          % Here are the pacs numbers
          % Here are the pacs references
05.40.+j, % Fluctuation phenomena, random processes, and Brownian motion
07.79.Lh, % Atomic force microscopes
42.50.Lc, % Quantum fluctuations, quantum noise, and quantum jumps
39.90.+d  % Other instrumentation and techniques for atomic and molecular physics}
}
\vspace{-0.14in} % hack to buy space
\narrowtext
The Casimir force is the mean force between two objects that is 
generated by quantum fluctuations \cite{Landau:80}.  In a recent 
review article \cite{Barton:94}, Barton notes: ``It is strange that 
for nearly half a century after Casimir no curiosity has been 
displayed regarding the fluctuations [of the Casimir force] about the 
mean.''  This lack of curiosity is understandable in view of the 
prevailing opinion, as summarized in \cite{Barton:94}, that 
Casimir fluctuations are ``far too small to detect with any 
traditionally contemplated Casimir-type apparatus.''

In this article we venture a contrary prediction---that Casimir force 
fluctuations are large enough to be directly detected by force 
microscopes, and that these fluctuations provide a fundamental limit 
to force microscope sensitivity which is stringent enough to be 
significant in practical applications \cite{Sidles:95}.  We obtain 
this prediction via a strategy advocated by Kupiszewska 
\cite{Kupiszewska:92}:
\begin{quote} 
The standard macroscopic quantum theory for nonhomogenous 
media\,{\ldots}refers to a medium described by a constant refractive 
index.  Although useful for many applications, this approach, as well 
as all other approaches neglecting losses, is generally incorrect.  It 
is well known that the dielectric function must satisfy the 
Kramers-Kronig relations, otherwise causality would be violated.  
According to Kramers-Kronig relations, the imaginary part of a 
realistic, frequency-dependent dielectric function must not vanish, 
and that implies the dissipation of radiation energy.  Therefore, a 
complete theory will have to include not only the field and the atoms, 
but also a system that absorbs energy, usually called a heat bath or 
reservoir.
\end{quote}
In implementing Kupiszewska's program, we will consider the two force 
microscope geometries shown in Fig.~\ref{fig:RayCantilever}.  Both 
geometries assume a spherical cantilever tip.  We confine our 
attention to experiments conducted \emph{in vacuo} at cryogenic 
temperatures \cite{Sidles:95}, because such experiments have the 
force sensitivity required to directly 
observe the predicted Casimir fluctuations.

Our discussion centers upon four parameters which change when the 
cantilever tip approaches the sample---our goal is to 
predict these changes.  The four parameters are: (a)~the resonant 
frequency $\omega_{0}$, (b)~the spring constant $k$, (c)~the 
cantilever resonant quality $Q$, and (d)~the force noise spectral 
density $S_{f}$.

\bfi[t]
{
    \vspace*{1.3in} % Extra vspace to accommodate our abstract hack
    \begin{center}
    \epsfxsize=3.3in
    \epsfbox{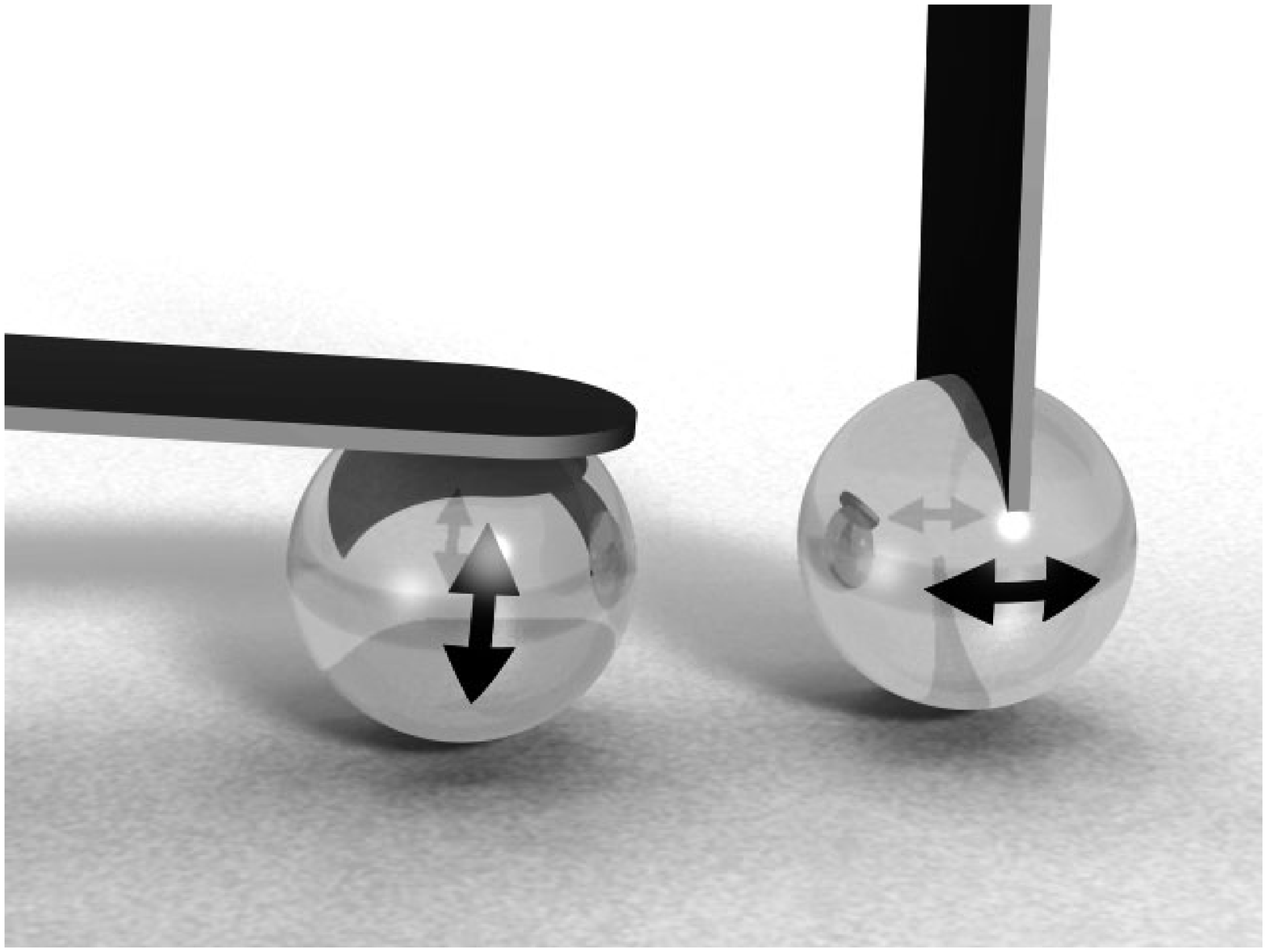} 
    \end{center}
}
\caption{Two force microscope geometries: at left, tip vibration normal 
to the sample plane; at right, tip vibration in the sample 
plane.\protect\vspace{-0.2in}}
\label{fig:RayCantilever} 
\efi 

We begin by describing how these parameters are measured.  In typical 
experiments \cite{Sidles:95}, the tip is brought near the sample and 
the Brownian motion $x(t)$ of the cantilever tip is observed 
interferometrically.  In well-designed experiments $x(t)$ is dominated 
by thermal noise, such that the cantilever motion is in thermal 
equilibrium with $k\langle{x^{2}}\rangle = k_{B}T$, where 
$\langle{x^{2}}\rangle$ is the mean square tip displacement, $k_{B}$ 
is Boltzman's constant, and $T$ is the ambient temperature.  In such 
experiments the autocorrelation of $x(t)$ is exponential:
\be 
{\langle{x(t)x(t+\tau)}\rangle}_{t} 
    = \langle x^{2}\rangle\ e^{-\omega_{0} \tau/(2 Q)}\ 
    \cos(\omega_{0}\tau),
        \label{eq:equilibrium}
\ee 
where $\langle\,\rangle_{t}$ denotes a time average.  The  
autocorrelation thus determines the parameters 
${\langle{x^{2}}\rangle}$, $\omega_{0}$, and $Q$.  From them, it is 
routine practice to: (a)~infer the spring constant via $k = m 
\omega_{0}^2$, with $m$ the motional mass of the cantilever,  
(b)~verify that thermodynamic equilibrium is respected by checking that 
$k {\langle{x^{2}}\rangle} = k_{B}T$, and (c)~calculate a Langevin 
force spectral density $S_{f}$ via
\bea 
    S_{f} & \equiv & \int_{-\infty}^{\infty}\!\!d\tau\ 
        e^{-i \omega \tau} \langle{f(t) f(t+\tau)}\rangle
        \nonumber\\ 
    & = & \frac{2 k^{2}}{Q \omega_{0}}\ \langle{x^{2}}\rangle = 
        \Big(\frac{m \omega_{0}}{Q}\Big)\ 2 k_{B}T.
        \label{eq:fluctuation}
\eea 
The term $m \omega_{0}/Q$ in (\ref{eq:fluctuation}) is recognizably 
the damping coefficient of a mechanical oscillator; (\ref{eq:fluctuation}) 
thus represents the fluctuation-dissipation theorem \cite{Balescu:75} 
as it applies to force microscope experiments.
% with $\hbar \omega_{0} \ll k_{B}T$ (an inequality which we henceforth 
% assume).
By this method, or inconsequential variants thereof, the four 
parameters $\omega_{0}$, $Q$, $k$, and $S_{f}$ are routinely measured 
in force microscopy.

For tip vibration normal to the sample plane (as in 
Fig.~\ref{fig:RayCantilever} at left), the spring constant $k$ 
decreases as the tip approaches the sample.  Our main prediction, 
which is derived in the second half of this article, is that the 
observed change in spring constant $\delta k$ will be accompanied by 
an increase in force noise $\delta S_{f}$ according to
\be 
    \delta S_{f} = \frac{9 \pi}{40 \ln(4/e)}\ \hbar\ (-\delta k).
        \label{eq:normalNoise} 
\ee
Then from (\ref{eq:fluctuation}) and (\ref{eq:normalNoise}), the 
dynamical cantilever damping $m \omega_{0}/Q$ is predicted to increase 
according to
\be
    \delta\Big(\frac{m \omega_{0}}{Q}\Big) = \frac{1}{2 k_{b}T}\ 
        \delta S_{f} = \frac{9 \pi}{80 \ln(4/e)}\ \frac{\hbar}{k_{B}T}\ (-\delta k),
        \label{eq:normalDamping}
\ee
thus ensuring that the Casimir fluctuations respect thermodynamic 
equilibrium, such that $k\langle{x^{2}}\rangle = k_{B}T$ for all 
values of $\delta k$.  Note that these predictions involve only 
experimentally measured quantities and Planck's constant---there are 
no model-dependent parameters.

Are the predicted force fluctuations large enough to observe directly?  
From (\ref{eq:normalNoise}), we compute that a spring constant shift 
$\delta k = -2.6\times 10^{-3}\,\text{N/m}$ will be associated with a 
temperature-independent Casimir force noise of 
$10^{-18}\,\text{N}/\sqrt{\text{\small Hz}}$.  Spring constant shifts 
of this magnitude are commonly observed in force microscopy.  Recent 
experiments have demonstrated force noise levels of order 
$10^{-17}\,\text{N}/\sqrt{\text{\small Hz}}$, and if force sensitivity 
continues to improve \cite{Sidles:95} then it is reasonable to expect 
that the predicted Casimir fluctuations will become a dominant noise 
mechanism in future force microscope experiments.

To describe experiments in which tip vibration is in the plane of the 
sample---as in Fig.~\ref{fig:RayCantilever} at right---we introduce a length 
scale $l$ defined such that an attractive Casimir force $f$ exerted 
between tip and sample generates a change in spring constant $\delta k 
= f/l$.  To calculate $l$ explicitly, let $\phi(z)$ be the modal 
eigenfunction of the cantilever, with $z$ the coordinate along the 
cantilever length $L$, normalized such that $\phi(L) = 1$.  Then $l$ 
is given by 
\be
    l^{-1} = \int_{0}^{L}\!\!dx\ 
        \left[\frac{\partial \phi(x)}{\partial x}\right]^2\!\!\!.
        \label{eq:modShape}
\ee
Typically $l/L$ is of order unity.  
% THE FOLLOWING TEXT IS OMITTED WITH PAIN
% In general $l$ must be computed by numerical integration, but for the 
% special case of a cantilever with uniform cross section and negligible 
% tip mass it can be shown analytically that the fundamental mode has $l 
% = 0.861L$.
Letting $h$ be the tip-to-sample separation distance, the predicted 
increase in force noise as the tip approaches the sample is
\be
      \delta S_{f} = 
          \frac{3 \pi}{160 \ln(4/e)}\ \frac{l}{h}\ \hbar\,\delta k.
          \label{eq:transverseNoise}
\ee
There is no minus sign in this equation, in contrast 
to~(\ref{eq:normalNoise}), because the end-on geometry shown at right 
in Fig.~\ref{fig:RayCantilever} yields a positive $\delta k$ at close 
tip-sample separation.  Then (\ref{eq:fluctuation}) and 
(\ref{eq:transverseNoise}) yield the predicted damping increase:
\be
    \delta\Big(\frac{m \omega_{0}}{Q}\Big) 
        = \frac{1}{2 k_{b}T}\ \delta S_{f} 
        = \frac{3 \pi}{320 \ln(4/e)}\ \frac{\hbar}{k_{B}T}\ \frac{l}{h}\ \delta k.
        \label{eq:transverseDamping}
\ee
Assuming $\delta k$ is independent of temperature to leading 
order---which is a reasonable assumption for Casimir forces---the 
predicted cantilever damping varies \emph{inversely} with temperature, 
according to (\ref{eq:normalDamping}) and 
(\ref{eq:transverseDamping}).  Such inverse relations are uncommon in 
physics but they are not unknown; the Kondo effect is an example.

It remains only to derive (\ref{eq:normalNoise}) and 
(\ref{eq:transverseNoise}) by the program of Kupiszewska.  We will not 
hesitate to make brutal simplifying approximations along the way, with 
a view toward obtaining results in closed form.  
Furthermore, we will finesse various model-dependent parameters by 
showing that they appear in $\delta S_{f}$ and $\delta k$ in such a 
manner that the ratio $\delta S_{f}/(\delta k)$ is 
parameter-independent.  By this strategy we can reasonably hope to 
obtain final results which have broader validity than the underlying 
model from which they derive.

Following Kupiszewska \cite{Kupiszewska:92}, we model individual atoms 
as independent harmonic oscillators.  By an appropriate scaling of 
variables the interaction of a tip atom $a$ with a sample atom $b$ can 
be described by the Hamiltonian
\bea
    H & = & \frac{1}{2}\,\omega_{a}\,(p_{a}^{2} + q_{a}^{2}) +  
        \frac{1}{2}\,\omega_{b}\,(p_{b}^{2} + q_{b}^{2}) 
        \nonumber\\
     & & -\ \beta_{ab} q_{a}q_{b} + (\text{heat bath}).
        \label{eq:Hamiltonian}
\eea
Here $\omega_{a}$ and $\omega_{b}$ are atomic oscillator frequencies 
and $\beta_{ab}$ is a dipole coupling whose strength and 
spatial dependence are discussed later.  The operators 
$\{q_{a},p_{a},q_{b},p_{b}\}$ obey the usual commutation relations 
$[q_{i},p_{j}]\!=\!i\hbar\,\delta_{ij}$, and it will turn out that no 
other information need be specified about their physical nature.  
We specify the heat bath as the unique independent oscillator (IO) 
model of Ford \emph{et al.} \cite{Ford:88} that induces 
linear damping in the Heisenberg picture equations of motion:
\bea
    {\ddot q}_{a}+\Gamma_{\!a}\omega_{\!a}
        {\dot q}_{a} + \omega_{\!a}^{2} q_{a} 
        & = & F_{a}(t) + \beta_{ab} q_{b}
        \label{eq:EOM1}\\
    {\ddot q}_{b} + \Gamma_{b}\omega_{b}
        {\dot q}_{b} + \omega_{b}^{2} q_{b}
        & = & F_{b}(t) + \beta_{ab} q_{a}.
        \label{eq:EOM2}
\eea
Here $F_{a}(t)$ and $F_{b}(t)$ are operator-valued Langevin forces 
which originate in the heat bath, and $\{\Gamma_{a},\Gamma_{b}\}$ are 
damping rates.  We pause to note that our conventions 
for correlation and power spectral density are those of Balescu's 
textbook \cite{Balescu:75}:
\bea
    C_{AB}(\tau) &\equiv& 
        \frac{1}{2}\ \langle{A(t)B(t+\tau)+B(t+\tau)A(t)}\rangle
        \nonumber\\
    S_{A}(\omega) &\equiv& 
        \int_{-\infty}^{\infty}\!\!d\tau\ e^{i 
        \omega \tau} C_{AA}(\tau),
        \nonumber
\eea
with $\langle{\,}\rangle$ an expectation over heat bath variables.  
Then as shown by Ford \emph{et al.} \cite{Ford:88}, the assumption of 
linear damping in (\ref{eq:EOM1}-\ref{eq:EOM2}) uniquely determines 
the Langevin force autocorrelation to be
\be
    C_{F_{i}F_{j}}(\tau) = 
        \delta_{ij}\,\frac{\Gamma_{i}\omega_{i}}{2 \pi}\!\!
        \int_{-\infty}^{\infty}\!\!\!d\omega\,\hbar \omega\,
        \coth\!\big(\frac{\hbar \omega}{2 k_{B}T}\big)\,e^{-i\omega\tau}.
        \label{eq:EOM3}
\ee The resulting model of atomic fluctuations resembles Kupiszewska's 
model of gauge field fluctuations quite closely.  Kupiszewska's 
model integrates over matter fields to obtain equations in which only 
gauge fields appear.  Our strategy is the opposite: we have integrated 
over the longitudinal gauge fields which generate the atomic dipole 
coupling \cite{Bjorken:65}, such that (\ref{eq:EOM1}--\ref{eq:EOM3}) 
contain only matter fields.  The two approaches are formally 
equivalent, because in the real world gauge fluctuations and matter 
fluctuations are inseparable, such that either can be regarded as the 
fundamental dynamical variable.

% ANOTHER PAINFUL OMISSION
% However, our respective models serve two very different purposes.  
% Kupiszewska's model is a vehicle for studying dispersive 
% corrections to field-theoretic models---it is 
% one-dimensional, encompasses only transverse photons, and does not 
% treat force fluctuations.  In contrast, our atomic model is 
% three-dimensional, encompasses only longitudinal photons---which 
% dominate at close range---and is specifically focused on force 
% fluctuations.
% 
We now wish to compute the pairwise Casimir force ${\bf f}_{ab}$ 
between two atoms, the associated spring constant ${\bf k}_{ab}$, and 
the force spectral density ${\bf S}_{{\bf f}_{ab}}(\omega)$.  The 
force is given in terms of the gradient of the dipole coupling by 
\cite{Balescu:75}
\be 
    ({\bf f}_{ab})_{i} = 
        ({\nabla}_{i}\beta_{ab})\,\langle{q_{a}q_{b}}\rangle, 
        \label{eq:forceDefinition} 
\ee
from which ${\bf k}_{ab}$ and ${\bf S}_{{\bf f}_{ab}}(\omega)$ follow 
immediately as
\bea 
    ({\bf k}_{ab})_{ij}& = &  
        -{\nabla}_{i}({\bf f}_{ab})_{j} 
       \label{eq:springDefinition} \\
    \big({\bf S}_{{\bf f}_{ab}}(\omega)\big)_{ij} & = & 
        ({\nabla}_{i} \beta_{ab})({\nabla}_{j} \beta_{ab})\ 
        S_{(q_{a}q_{b})}(\omega),\label{eq:noiseDefinition}
\eea
where the spectral density $S_{(q_{a}q_{b})}(\omega)$ is given by
\be
   S_{(q_{a}q_{b})}(\omega) =
        \int_{-\infty}^{\infty}\!\!\!\!d\tau\ 
        e^{i\omega \tau}\,C_{(q_{a}q_{b})(q_{a}q_{b})}(\tau).
        \label{eq:sqdefinition}
\ee
The only dynamical quantities that appear in 
(\ref{eq:forceDefinition}--\ref{eq:sqdefinition}) are the expectation 
$\langle{q_{a}q_{b}}\rangle$ and the autocorrelation 
$C_{(q_{a}q_{b})(q_{a}q_{b})}(\tau)$.  Our next step, therefore, is to 
compute these quantities to leading and next-to-leading order in 
$\{\beta_{ab},\Gamma_{a},\Gamma_{b}\}$, which are assumed small 
compared to $\{\omega_{a},\omega_{b}\}$.  Physically, the atomic 
oscillators are assumed to be weakly coupled by the Casimir 
interaction and underdamped by the ambient heat bath.

As described in \cite{Ford:88} and \cite{Li:93}, any desired 
correlation involving $q_{a}(t)$ and $q_{b}(t)$ can be explicitly 
computed by solving (\ref{eq:EOM1}--\ref{eq:EOM3}) in the Fourier 
domain.  For $\langle{q_{a}q_{b}}\rangle$ an uncomplicated Fourier 
integration yields
\bea
    \langle{q_{a}q_{b}}\rangle & 
        \stackrel{\scriptscriptstyle{T \rightarrow 0}}{\displaystyle{=}} & 
        \frac{\hbar \beta_{ab}}{2 (\omega_{a}+\omega_{b})} + 
        \Big[ \frac{\hbar \beta_{ab} \omega_{a} \omega_{b} }{2 \pi 
        (\omega_{a}^{2}-\omega_{b}^{2})^{2}} 
        \nonumber\\
     & &  \times \Big(\Gamma_{a} 
        \big(1-\frac{\omega_{a}^{2}}{\omega_{b}^{2}} + 
        \ln(\frac{\omega_{a}^{2}}{\omega_{b}^{2}})\big)+ 
        (a\!\leftrightarrow\!b)\Big)\Big],
        \label{eq:expectqaqb}
\eea
plus ${\cal O}(\beta^{3},\beta\,\Gamma^{2})$ terms.  The zero of the 
denominator in (\ref{eq:expectqaqb}) for $\omega_{a}\!=\!\omega_{b}$ 
cancels against a zero of the numerator, such that 
$\langle{q_{a}q_{b}}\rangle$ is finite for all nonzero $\omega_{a}$ 
and $\omega_{b}$.  Setting $\Gamma_{a}=\Gamma_{b}=0$ in 
(\ref{eq:expectqaqb}), we recover the same expression for 
$\langle{q_{a}q_{b}}\rangle$ as is obtained from the ground 
state of the Hamiltonian with the heat bath turned off.

These results establish that heat bath damping does not alter the 
interatomic Casimir force in leading order, as is physically 
reasonable.  

With regard to the autocorrelation 
$C_{(q_{a}q_{b})(q_{a}q_{b})}(\tau)$ in (\ref{eq:sqdefinition}), the 
Gaussian property of the Langevin forces, as discussed by Ford 
\emph{et al.} \cite{Ford:88}, allows it to be written as the product 
of two simpler autocorrelations:
\be
    C_{(q_{a}q_{b})(q_{a}q_{b})}(\tau) 
        = C_{q_{a}q_{a}}(\tau)\, C_{q_{b}q_{b}}
        (\tau) + {\cal O}(\beta_{ab}^{2}).\label{eq:factor}
\ee
In turn, $C_{q_{a}q_{a}}(\tau)$ is calculated by solving 
(\ref{eq:EOM1}--\ref{eq:EOM3}) in the Fourier domain in a manner that 
precisely parallels the calculations of Li \emph{et al.} \cite{Li:93}.  
The result is:
\bea
    C_{q_{a} q_{a}}(\tau) 
        & \stackrel{\scriptscriptstyle{T \rightarrow 0}}{\displaystyle{=}}&
        \frac{\hbar}{\pi}
        \int_{a}^{\infty}\!\!\!d\omega\,
        \frac{\Gamma_{a}\,\omega_{a}\,\omega\ \cos(\omega \tau)}
        {\left|\omega_{a}^{2}-\omega^{2}+i\Gamma_{a}\omega\right|^{2}} 
        + {\cal O}(\beta_{ab}^{2})
        \label{eq:integral}\\
    &=& \frac{\hbar}{2}\ \frac{\omega_{a}}{{\bar\omega}_{a}}\ 
        \Big[e^{-\Gamma_{a} |\tau|/2}\,\cos({\bar\omega}_{a}\tau)
        \nonumber \\
    & & +\ \frac{2}{\pi}\,\text{Im}\,
        g\big(({\bar\omega}_{a} + \frac{i\Gamma_{a}}{2}) |\tau|\big)\Big]
        +{\cal O}(\beta_{ab}^{2})
        \label{eq:fullForm}\\
    &=& \frac{\hbar}{2}e^{-\Gamma_{a} |\tau|/2}\cos({\bar\omega}_{a}\tau)
        + {\cal O}(\Gamma_{a},\beta_{ab}^{2}).
        \label{eq:correlation}
\eea
By substituting $a\rightarrow b$ we obtain $C_{q_{b} q_{b}}(\tau)$.  
Here ${\bar\omega}_{a} \equiv (\omega_{a}^{2}-\Gamma_{a}^{2}/4)^{1/2}$ 
is assumed real and positive, and $g(z)$ is the exponential integral 
\cite{Abramowitz:65} defined by
\be
    g(z) \equiv \int_{0}^{\infty}\!\!dt\ \frac{\cos(t)}{t+z}.
        \label{eq:exponential}
\ee
The term $\text{Im}\,g\big(({\bar\omega}_{a} + i \Gamma_{a}/2) 
|\tau|\big)$ in (\ref{eq:fullForm}) can be proved to be monotonic in 
$|\tau|$, with initial value $-\tan^{-1}(\Gamma_{a}/(2 
{\bar\omega}_{a}))$ and asymptotic value $-\Gamma_{a} 
{\bar\omega}_{a}/(\tau^{2}\omega_{a}^{4})$.  This term therefore 
describes squeezing of the quantum zero-point motion by the heat bath 
damping---a phenomenon which is physically to be expected.  

% Numerical 
% comparison of (\ref{eq:fullForm}) and (\ref{eq:correlation}) 
% establishes that squeezing is a reasonably small correction for all 
% $\Gamma_{a} \alt \omega_{a}$.

The engineering import of~(\ref{eq:correlation}) is that even at zero 
temperature, where classical oscillators exhibit zero noise, the 
zero-point motion of a lightly damped ($\Gamma_{a}\!\ll\! \omega_{a}$)
quantum oscillator carries $\hbar/2$ of noise power within a bandwidth 
$\Gamma_{a}$ centered on a carrier frequency $\omega_{a}$.

Since each atom in our model is coupled to an independent heat bath, 
and is dynamically independent of adjacent atoms, we can obtain the 
total Casimir force and force noise by summing the atomic interactions 
pairwise.  For definiteness, we assume a Debye distribution of atomic 
frequencies $\{\omega_{a},\omega_{b}\}$, such that $p(\omega) = 3 
\omega^{2}/\omega_{D}^{3}$, where $\omega_{D}$ is a Debye 
frequency---any other broad-band frequency distribution would yield 
similar results.  Then we frequency-average 
$\langle{q_{a}q_{b}}\rangle$ from (\ref{eq:expectqaqb}) and 
$S_{(q_{a}q_{b})}(\omega)$ from (\ref{eq:sqdefinition}), 
(\ref{eq:factor}), and (\ref{eq:correlation}) to obtain to leading 
order in $\{\beta_{ab},\Gamma_{a},\Gamma_{b}\}$:
\bea
    \langle{q_{a}q_{b}}\rangle^{\text{\it av}} & = & 
        \int_{0}^{\omega_{D}}\!\!\!\!\!d\omega_{a}\,d\omega_{b}\ 
        p(\omega_{a})p(\omega_{b})\ 
        \langle{q_{a}q_{b}}\rangle\nonumber\\
        & = & \frac{9 \ln(4/e)}{10}\ 
        \frac{\hbar \beta_{ab} }{\omega_{D}}\label{eq:theqaqb}\\
	S_{(q_{a}q_{b})}^{\text{\it av}}(\omega) &=& 
        \int_{-\infty}^{\infty}\!\!\!\!d\tau\!\!
        \int_{0}^{\omega_{D}}\!\!\!\!\!d\omega_{a}\,d\omega_{b}\,
        \Big[ p(\omega_{a})\,p(\omega_{b})\,e^{i\omega\tau}\nonumber\\
       && \times\ 
       C_{(q_{a}q_{b})(q_{a}q_{b})}(\tau)\Big]  
    \stackrel{\scriptscriptstyle{\omega \rightarrow 0}}{\displaystyle{=}} 
       \frac{9 \pi \hbar^{2}}{20 \omega_{D}}
        \label{eq:theotherkey}
\eea
In (\ref{eq:theotherkey}) we have assumed $\omega \ll \omega_{D}$, as 
appropriate for audio frequency force microscope experiments.

It remains only to integrate over the tip and sample volumes shown in 
Fig.~\ref{fig:RayCantilever}.  We specify that the tip and sample 
atoms have number density $\rho_{a}$ and $\rho_{b}$ respectively.  
The coupling $\beta_{ab}$ is assumed to have a dipole dependence 
$\beta_{ab} = \kappa/|{\bf r}_{ab}|^{3}$, with ${\bf r}_{ab}$ the 
atomic separation vector and $\kappa$ the strength of the dipole 
interaction.  Substituting (\ref{eq:theqaqb}--\ref{eq:theotherkey}) 
into (\ref{eq:forceDefinition}--\ref{eq:noiseDefinition}) and 
integrating over the volume of a spherical tip of radius $r$ that is 
separated by a gap $h$ from a half-space sample, we obtain the total 
Casimir force, spring constant, and force noise as
\bea 
    {\bf f}^{\text{\it{tot}}} &=& 
        -{\bf \hat n}\ \frac{\hbar \rho_{a} \rho_{b} 
        \kappa^{2}}{\omega_{D}}\ 
        \frac{3 \ln(4/e) \pi^{2}}{10}\ 
        \frac{r^{3}}{h^{2} (2r+h)^{2}}\label{eq:ftot}
        \\
    {\bf k}^{\text{\it{tot}}} & = & 
        -({\bf \hat n}\otimes{\bf \hat n})\ 
        \frac{\hbar \rho_{a} \rho_{b} \kappa^{2}}{\omega_{D}}\nonumber\\
        && \times
       \frac{6 \ln(4/e) \pi^{2} }{5}\ 
        \frac{r^{3}(r+h)}{h^{3} (2r+h)^{3}}
        \\
    {\bf S}_{\bf f}^{\text{\it{tot}}} & = & 
        \Big[{\bf \hat n}\otimes{\bf \hat n}
        + \frac{1}{24} (\text{\bf I}-{\bf \hat n}\otimes{\bf \hat n})\Big]
        \nonumber\\
     & & \times\,\frac{\hbar^{2}\rho_{a} \rho_{b} \kappa^{2}}{\omega_{D}}\ 
        \frac{27\pi^{3}}{100}\ 
        \frac{r^{3}(r+h)}{h^{3} (2r+h)^{3}}.
        \label{eq:forceNoiseTensor}
\eea 
Here $({\bf I})_{ij}\equiv\delta_{ij}$ is the identity matrix, ${\bf 
\hat n}$ is a unit vector normal to the sample surface, and $({\bf 
\hat n}\otimes{\bf \hat n})_{ij} \equiv {\bf \hat n}_{i}{\bf \hat 
n}_{j}$ is an outer product.
% ANOTHER PAINFUL OMISSION
% It can be shown that 
% (\ref{eq:ftot}) corresponds to a Casimir force per unit area 
% proportional to $1/h^{3}$, as expected for nonretarded interactions.  
% We note in (\ref{eq:forceNoiseTensor}) that vertical force 
% fluctuations carry precisely 24 times the noise power of transverse 
% force fluctuations.  
With the neglect of ${\cal O}(h/r)$ terms, as is reasonable for close 
tip-sample approach, our main results (\ref{eq:normalNoise}) and 
(\ref{eq:transverseNoise}) follow immediately from 
(\ref{eq:ftot}--\ref{eq:forceNoiseTensor}).

Consistent with our stated goal of model independence, we need not 
specify numerical values for the parameters 
$\{\rho_{a}$,$\rho_{b}$,$\kappa$,$\omega_{D},\Gamma_{a},\Gamma_{b},h,r\}$, 
because they do not appear in the final results.  However, the 
dimensionless coefficients in (\ref{eq:normalNoise}) and 
(\ref{eq:transverseNoise}) are weakly sensitive to the functional form
of the assumed Debye distribution of atomic frequencies.  Thus, for 
example, the coefficient $9 \pi/(40 \ln(4/e)) \sim 1.83$ appearing in 
(\ref{eq:normalNoise}) is best regarded as a coefficient of order 
unity, whose precise value will depend on the material properties and 
shape of the tip and sample.  These coefficients are well suited for
experimental determination.

Arguably the two least realistic assumptions of our model are the 
assumed dynamical independence of adjacent atoms, and the coupling of 
each atom to an independent heat bath possessed of an infinite number 
of degrees of freedom.  The path to a more realistic model is clear 
but arduous.  The heat bath model should be improved to describe 
realistic phonon and conduction band degrees of freedom, while taking 
into account the finite size of the tip, and electronic degrees of 
freedom in adjacent atoms should be realistically coupled.  Both gauge 
fields and matter fields should be explicitly included in the 
Hamiltonian as in Kupiszewska's pioneering article 
\cite{Kupiszewska:92}.  The resulting field equations should be solved 
for realistic tip-sample geometries.  Force fluctuations should be 
computed by the field-theoretic methods pioneered by Barton 
\cite{Barton:91,Barton:91a}.  Ideally, the results should explicitly 
respect the fluctuation-dissipation theorem and should be expressed in 
a simple and physically transparent form.

Meeting these challenges will not be easy.  Yet if the predictions of 
this article are experimentally confirmed, such that Casimir effects 
set the practical limits to force microscope sensitivity, then 
achieving a realistic understanding of these effects will become a 
matter of urgent practical consequence, in particular to the 
biomedical research community \cite{Sidles:95}.  And if the 
predictions of this article are not confirmed, the question will 
be:~why~not?

In either case, it is certain that Casimir effects will continue to 
engage and delight the imaginations of the physicists and engineers in 
coming decades.
\vspace{-0.15in}

\acknowledgements{
\protect\vspace{-0.15in}
This work was supported by the NIH Biomedical Research Technology 
Program and the NSF Major Research Instrument (MRI) Program.}

% \bibliographystyle{prsty}
% \bibliography{Casimir}

\vspace{-0.2in} % vspace to fit on four pages

\end{document}